\documentclass[twocolumn,pre,showpacs]{revtex4-1}

\usepackage{graphicx}
\usepackage{amsmath}
\usepackage{amssymb}

\begin{document}

\title{Comment on "Phase transition in a one-dimensional Ising ferromagnet
	at zero temperature using Glauber dynamics with a synchronous updating mode"}
\author{Il Gu Yi}
\author{Beom Jun Kim}
\email[Corresponding author, E-mail: ]{beomjun@skku.edu}
\affiliation{BK21 Physics Research Division and Department of Physics, Sungkyunkwan University, Suwon 440-746, Korea}


\begin{abstract}
Sznajd-Weron in [Phys. Rev. E {\bf 82}, 031120 (2010)] suggested that the
one-dimensional Ising model subject to the zero temperature
synchronous Glauber dynamics exhibits a discontinuous phase transition.
We show here instead that the phase transition is of a continuous nature
and identify critical exponents: $\beta \approx 0$, $\nu \approx 1$, and $z \approx 2$, 
via a systematic finite-size scaling analysis.
\end{abstract}



\maketitle

Recently, Sznajd-Weron~\cite{SW} has studied the phase transition in a 
one-dimensional (1D) 
Ising ferromagnet at zero temperature using Glauber dynamics
with a synchronous update rule. Interestingly, it has been successfully
shown that the system exhibits a well-defined phase transition as the
parameter $W_0$, the spin flipping probability for the null energy difference, 
is changed.
Whereas the absence of the phase transition at a finite temperature 
in the standard equilibrium Ising chain model is a well-known textbook
example, the slow magnetic relaxation in the spin chain systems has become
a hot research topic, theoretically and experimentally (see references in~\cite{SW}).

In the Glauber dynamics of the 1D ferromagnetic spin chain at zero
temperature, the energy difference $\Delta E$ is computed for the single-spin
flipped configuration, and the spin flip is accepted at the
probability~\cite{SW} 
\begin{equation}
W(\Delta E) = \left\{ \begin{array}{ll}
		1, & \textrm{for} \quad \Delta E < 0, \\
		W_{0}, & \textrm{for} \quad \Delta E = 0, \\
		0, & \textrm{for} \quad \Delta E > 0.
		\end{array}  \right.
	\label{eq:glauber}
\end{equation}
For each site of the system, the above probability is computed and each
spin flip at next time step is decided. The synchronous update rule 
means that all spins in the system are updated at the
same time, differently from more commonly used method of the sequential spin
update rule. The use of the synchronous update rule enables the system
to have the antiferromagnetic ordering, different from the use of the
sequential update rule~\cite{SW}.
As an order parameter, the density $\rho$ of active bonds which connect
different spin values (up and down) is used:
\begin{equation}
\rho = \frac{1}{2L} \sum_{i=1}^{L} (1- \sigma_{i} \sigma_{i+1}), 
\end{equation}
where $L$ is the number of spins and $\sigma_i(=\pm 1)$ is the Ising
spin at the $i$th site in the 1D chain with the periodic boundary condition
$\sigma_{i+L} = \sigma_i$ applied. The density of active bonds
is especially useful in the  present context, since it can effectively
distinguish the ferromagnetic ordering ($\rho = 0$) and the antiferromagnetic
ordering ($\rho = 1$). It should be noted that the system eventually
approaches the steady state which is either fully 
ferromagnetic $\rho_{st} = 0$ or fully antiferromagnetic $\rho_{st} = 1$,
and that the intermediate value inbetween is not possible as the time 
$t \rightarrow \infty$. Due to the zero temperature nature of the dynamics,
the ergodicity is broken and thus the time average and the ensemble average
of $\rho_{st}$ are not equivalent to each other. 
The ensemble average of the order parameter $\langle \rho_{st}
\rangle$ in the present context equals the probability of the fully
antiferromagnetic stationary state, and thus it can change continuously
across the phase transition at $W_c$. We emphasize that the {\it continuity} of
$\langle \rho_{st} \rangle$ does not contradict the {\it discreteness} of $\rho_{st}$.


\begin{figure}
\includegraphics[width=0.42\textwidth]{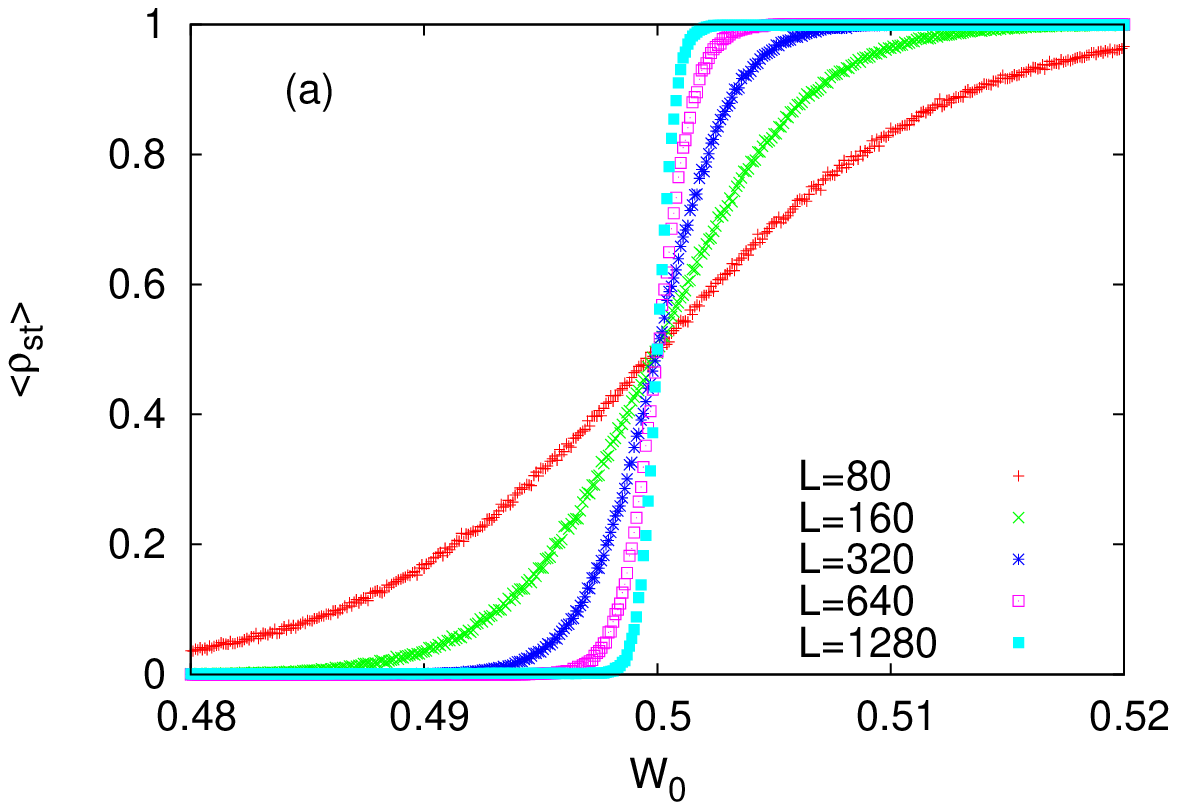}
\includegraphics[width=0.42\textwidth]{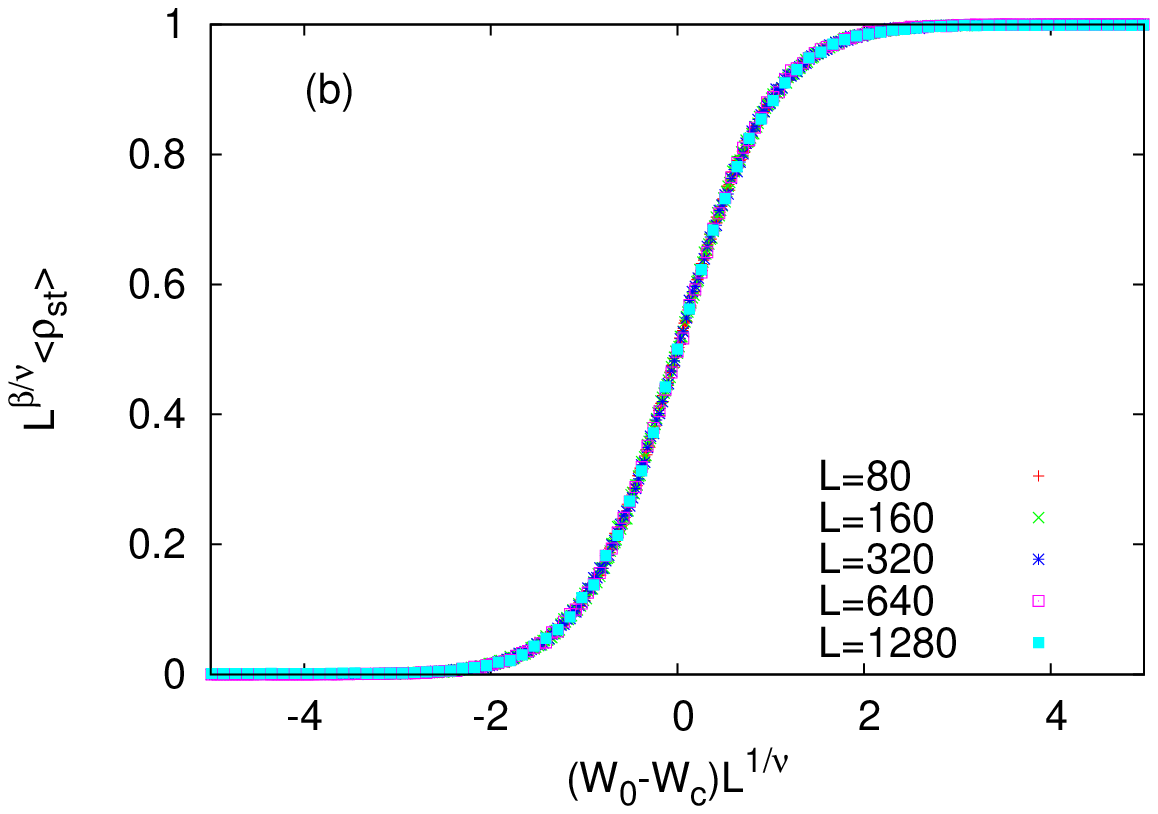}
\caption{(Color online) (a) Ensemble average $\langle \rho_{st} \rangle$ of the
density of active bonds in the steady state versus the spin flipping probability
$W_0$ for null energy difference, at various system sizes $L=80, 160, 320,
640$, and $1280$. (b) Finite-size scaling of data points in (a) using the
scaling form in Eq.~(\ref{eq:rhoscaling}) with $W_c = 0.5$, $\beta =0$, and
$\nu = 1$. The existence of a {\em continuous} phase transition is clearly
shown. Ensemble averages were performed over $10^{4}$ random initial spin
configurations. }
\label{fig:rho}
\end{figure}

Figure~\ref{fig:rho}(a), corresponding to Fig.~4 in \cite{SW}, 
displays the ensemble-average of the steady state value of
the active bond density $\langle \rho_{st} \rangle$ as a function of
the spin flipping probability $W_0$ for the null energy difference. 
We believe that our simulation results and the ones presented in~\cite{SW}
are identical and both clearly indicate the existence of the phase
transition at $W_0 = W_c (\approx  0.5)$. However, in a sharp contrast to \cite{SW}
where a discontinuous phase transition has been concluded, 
in this Comment we claim that the system exhibits a continuous 
phase transition.
In order to support our claim of a continuous phase transition,
we apply the standard technique of the finite-size scaling~\cite{goldenfeld}
of the order parameter $\langle \rho_{st} \rangle$:
\begin{equation} \label{eq:rhoscaling}
\langle \rho_{st} \rangle = L^{-\beta/\nu} f\bigl( (W_{0} - W_{c})L^{1/\nu}  \bigr),
\end{equation}
where $f(x)$ is a universal scaling function, and $\beta$ and $\nu$ are
standard critical exponents, describing the criticality of the order
parameter and the coherence length, respectively.
In Fig.~\ref{fig:rho}(b), we show the scaling collapse of $\langle
\rho_{st}\rangle$ through the use of the scaling form~(\ref{eq:rhoscaling}).
It is very clearly shown that the phase transition at $W_c \approx 0.5$ 
is well captured by $\beta \approx  0.0$ and $\nu \approx 1.0$. 
The well-defined value of $\nu$ indicates that the coherence length $\xi$ 
in the system diverges as the critical point is approached, as in a usual 
continuous phase transition~\cite{foot}.

\begin{figure}
\includegraphics[width=0.42\textwidth]{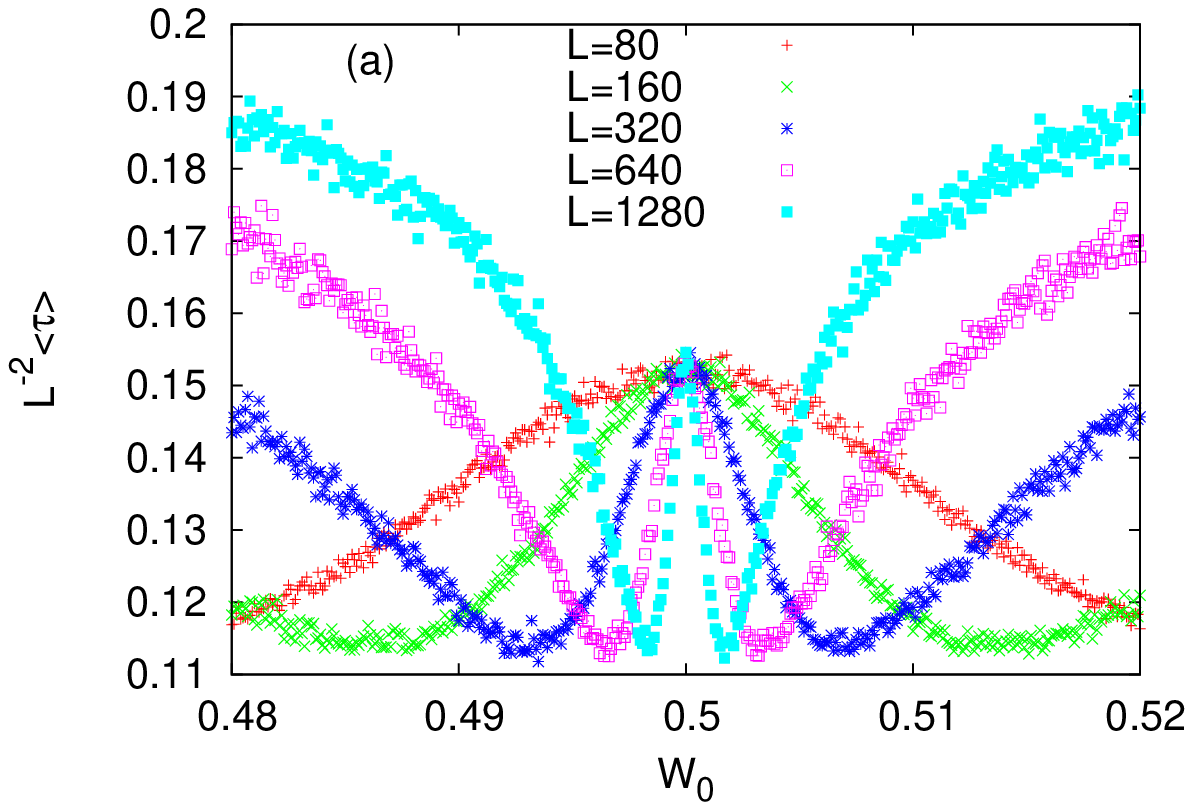}
\includegraphics[width=0.42\textwidth]{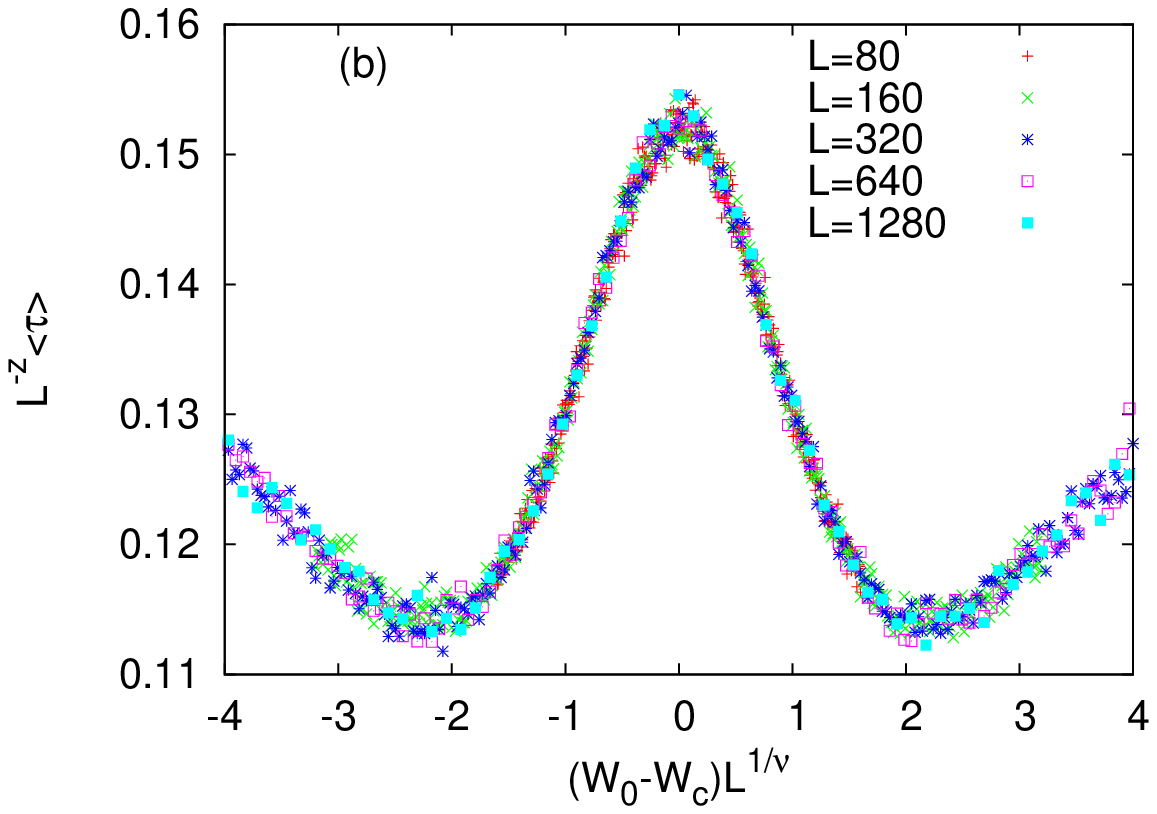}
\caption{(Color online) (a) The ensemble averaged relaxation time 
$\langle \tau \rangle$ multiplied by $L^{-2}$ 
versus the flipping probability $W_{0}$ for various system sizes.
(b) Finite-size scaling of relaxation time using the scaling form 
in Eq.~(\ref{eq:tauscaling}) with $W_c = 0.5$, $\nu =1$, and
$z = 2$. }
\label{fig:tau}
\end{figure}

The relaxation time $\tau$ for the system to reach the steady state ($\rho=0$
or $1$) from a random initial condition has also been reported in Fig.~5 of
\cite{SW}.  We again measure $\tau$ in the same way as in \cite{SW} 
and plot it in Fig.~\ref{fig:tau}(a).  As was already observed in~\cite{SW},
we also notice that $\langle \tau \rangle$ scales as $L^2$ at $W_c$. 
This reminds us that the dynamic critical exponent $z$~\cite{goldenfeld} 
describing 
the divergence of the relaxation time is given by $\tau \sim L^z$ 
exactly at the critical point. 
Accordingly, the observation of $\tau \sim L^2$ strongly indicates 
$z = 2$, and we can write the finite-size scaling form of the relaxation 
time as
\begin{equation}
\label{eq:tauscaling}
\langle \tau \rangle
= L^z g\bigl((W_{0} - W_{c}) L^{1/\nu}  \bigr),
\end{equation}
where $g(x)$ is the scaling function for the relaxation time. The finite-size
scaling form~(\ref{eq:tauscaling}) satisfies both 
$\langle \tau \rangle \sim L^z$ at $W_0 = W_c$ and $\langle \tau \rangle \sim \xi^z$
for $W_0 \neq W_c$ with the prescription $g(x) \sim x^{-\nu z}$ for small $x$
since the correlation length follows $\xi \sim (W_0 - W_c)^{-\nu}$.
In Fig.~\ref{fig:tau}(b), the scaling collapse of the relaxation time is
displayed, which confirms again the existence of the continuous phase transition
at $W_0 = W_c ( \approx 0.5)$, which is characterized by the coherence length
exponent $\nu \approx 1.0$ and the dynamic critical exponent $z \approx 2.0$.

In summary, different from the conclusion in~\cite{SW}, we have clearly shown
that the Ising ferromagnet chain at zero temperature with the synchronous
update rule exhibits a continuous phase transition at $W_0 = W_c (\approx 0.5)$.
Furthermore, through the use of the standard method of the finite-size scaling,
we have identified critical exponents $\beta \approx 0.0$, 
$\nu \approx  1.0$, and $z \approx 2.0$. 

This research was supported by Basic Science Research Program through the
National Research Foundation of Korea (NRF) funded by the Ministry of
Education, Science and Technology (2010-0008758).


\begin{thebibliography}{99}
\bibitem{foot}
Although not shown here, we have
confirmed that the magnetization and its Binder's cumulant, the staggered
magnetization and its Binder's cumulant, and also the Binder's cumulant 
for $\rho_{st}$,  all show good quality of finite-size scalings with the values
$W_c = 0.5$, $\beta = 0$, and $\nu = 1$.
\end{thebibliography}
\end{document}